# Evidence of Early Supershear Transition in the Mw 7.8 Kahramanmaraş Earthquake From Near-Field Records

Ares J Rosakis[1], Mohamed Abdelmeguid[1], Ahmed Elbanna[2,3]

[1] Graduate Aerospace Laboratories, California Institute of Technology, Pasadena, CA
[2] Department of Civil and Environmental Engineering, University of Illinois at Urbana Champaign, Urbana, Illinois
[3] Beckman Institute of Advanced Science and Technology, University of Illinois at Urbana Champaign, Urbana, Illinois

Abstract:
The $M_w 7.8$ Kahramanmaraş Earthquake was larger and more destructive than what had been expected for the tectonic setting in Southeastern Turkey. By using near-field records we provide evidence for early supershear transition on the splay fault that hosted the nucleation and early propagation of the first rupture that eventually transitioned into the East Anatolian fault. The two stations located furthest from the epicenter show a larger fault parallel particle velocity component relative to the fault normal particle velocity component; a unique signature of supershear ruptures that has been identified in theoretical and experimental models of intersonic rupture growth. The third station located closest to the epicenter, while mostly preserving the classical sub-Rayleigh characteristics, it also features a small supershear pulse clearly propagating ahead of the original sub-Rayleigh rupture. This record provides, for the first time ever, field observational evidence for the mechanism of intersonic transition. By using the two furthest stations we estimate the instantaneous supershear rupture propagation speed to be $$\sim 1.55 C_s$$ and the sub-Rayleigh to supershear transition length to be around $\sim 19.45$ km, very close to the location of the station nearest to the epicenter. This early supershear transition might have facilitated the continued propagation and triggering of slip on the nearby East Anatolian Fault leading to amplification of the hazard. The complex dynamics of the Kahramanmaraş earthquake warrants further studies.

## Introduction:

On February 6th 2023, a $M_w$ 7.8 earthquake shook the southeastern parts of Turkey and northern Syria. Preliminary back projection models based on teleseismic data as well as multiple seismic inversions suggest that rupture initiated at 1:17:355 coordinated universal time (UTC) on a splay branch fault in the near proximity of the East Anatolian fault [1]. The precise location of the hypocenter is currently uncertain. The preliminary hypocenter location was estimated by AFAD to be 37.288°N 37.042°E [2] with a depth of approximately 8 km. It was also estimated by the USGS to be 37.166°N 37.042°E $\pm$ 6.3 km (indicated by the red star marker in Figure 1) with a depth of approximately 18$\pm$3 km [1]. The rupture then propagated north east subsequently transferring to the East Anatolian fault and starting a sequence of seismic events. Furthermore, subsequent preliminary geodetic inversions confirmed the multi-segment nature of the $M_w$ 7.8 rupture. The sequence of events resulted in catastrophic levels of destruction with substantial humanitarian and financial losses. Based on historical records, the magnitude of the event and the total rupture length were both much larger than expected for such a tectonic setting in southern Turkey [3]. This together with the intensity of the measured ground shaking motivated us to investigate the nature of rupture initiation, propagation, as well as the possibility of early supershear transition.

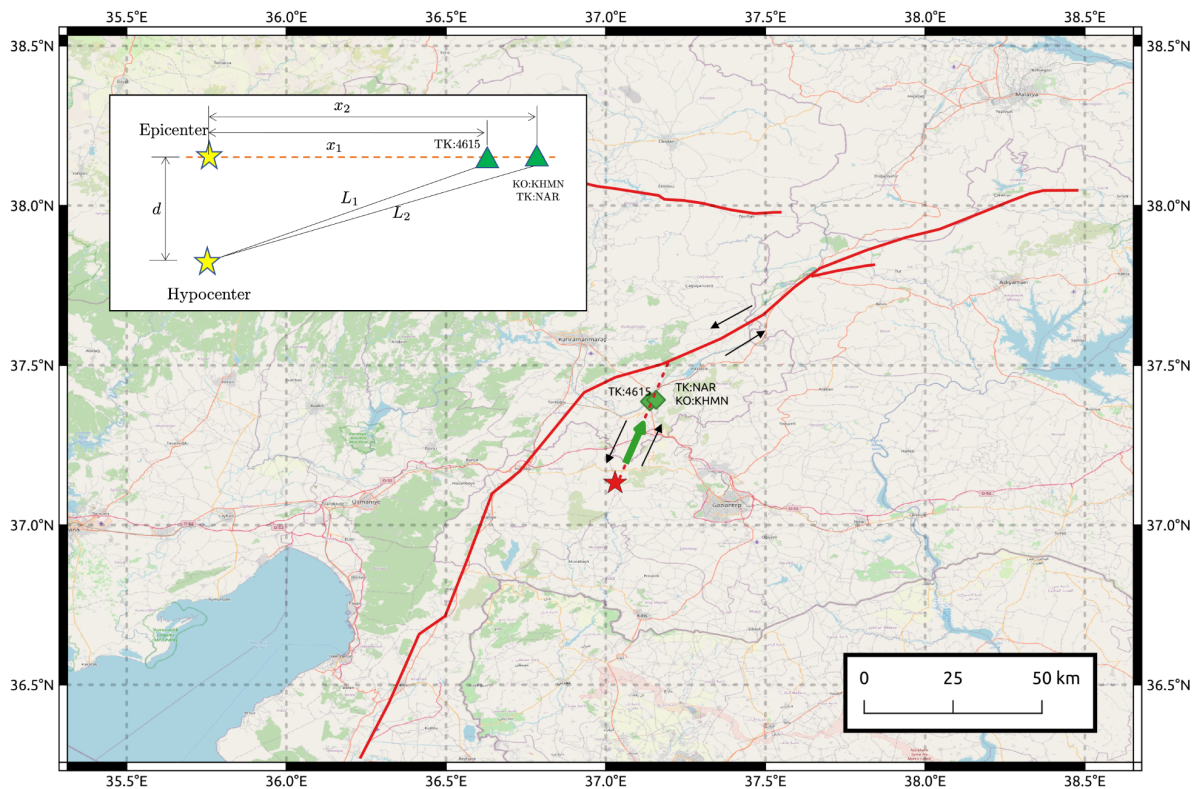

Figure 1. Map of the East Anatolian Fault (EAF) zone highlighting the estimated location of the hypocenter of the Mw7.8 Kahramanmaraş earthquake. The dashed line represents the inferred splay fault trace based on the recorded seismicity obtained

from AFAD. The green diamonds indicate the location of the nearest seismic station to the fault trace. The black arrows indicate the left lateral sense of motion of the fault. The insert is a schematic of the relative epicenteral and hypocentral locations of the stations.

Figure 1 illustrates the estimated location of the hypocenter, the approximate strike of the splay fault which is inferred to be around N22°E based on the aftershock sequence, and the sense of motion (left lateral for both the splay fault, and the east Anatolian fault). To the best of our knowledge, three stations exist very close to the splay fault as highlighted by the green diamonds in Figure 1. Two of these stations: TK:NAR and KO:KHMN are located at 37.3919°N 37.1574°E [2,4], and herein are referred to as the twin stations because they are at the same geographical location. Another station TK:4615 is located closer to the epicenter at 37.386°N 37.138°E [2]. The insert in Figure 1 is a schematic of the positions of the stations, showing the distances $x_1, x_2$ relative to the epicenter and the distances $L_1, L_2$ relative to the hypocenter which is located at a depth $d$. These three stations provide a rare and detailed insight into the near-field characteristics of the rupture on the splay fault and indeed close examination of these records have revealed unique observations that we describe below.

## Clear signature of supershear in the twin stations records

Figure 2a shows the time histories of the particle velocities along the fault parallel, the fault normal, and the vertical directions from the twin stations (TK:NAR solid black line, KO:KHMN solid red line). These are obtained from the instrument corrected ground motions. The raw NS, EW and vertical acceleration records are obtained from (AFAD) and (KOERI) respectively (Retrieved 02/09 5:18 PST) [2,4]. We computed the velocities for TK:NAR by numerically integrating the available acceleration records from AFAD [2]. The velocity response for KO:KHMN was processed using the Obspy software [5]. We then resolved the computed NS and EW ground velocity signals parallel and perpendicular to the splay fault shown in Figure 1. To the best of our knowledge, these records correspond to two different instruments and as a result the good agreement between the records provides a degree of confidence in the quality of the data to be used in the present study. Here, the first vertical dashed line indicates the first arrival of P-waves from the hypocenter based on the rupture initiation at the USGS provided time 1:17:355 coordinated universal time (UTC) [1].

The velocity waveforms for the twin stations reveal unique characteristics. We first observe that the FP component is clearly more dominant than the FN component. This is atypical of sub-Rayleigh strike-slip earthquake ruptures which feature more dominant fault normal versus fault parallel velocity components. However, a dominant fault parallel component is a characteristic feature of supershear ruptures [6] in which the rupture speed exceeds the shear wave speed of crustal rock $C_s$. Such a behavior has been observed both in the laboratory [7,8] and the field [8,9,10,11], and has been also predicted by the theory [7,9,12]. This provides evidence for supershear rupture propagation towards the twin stations.

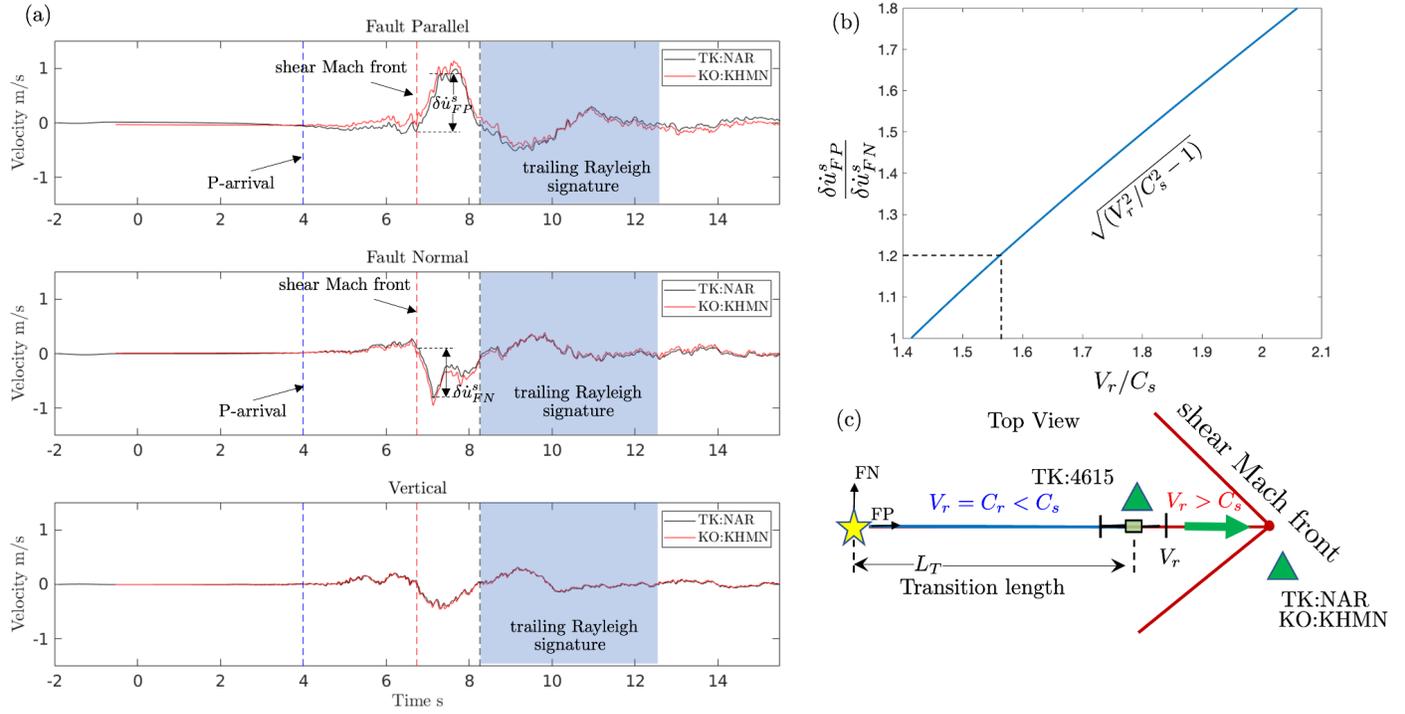

Figure 2. Supershear characteristics of near field records at stations TK:NAR, and KO:KHMN. (a) The instrument corrected records of the fault parallel, fault normal, and vertical particle velocities obtained at stations TK:NAR (black solid line), and KO:KHMN (red solid line). Note that the fault parallel component is larger than the fault normal component suggesting supershear rupture propagation. The blue dashed line indicates the arrival of the P-wave, the red dashed line indicates the arrival of the shear Mach front, and the black dashed line indicates the arrival of the trailing Rayleigh signature. (b) The theoretical relationship between the ratios of FP and FN velocity changes due the passage of the Mach front and supershear rupture speed normalized by the shear wave speed. For a ratio of velocity changes $= 1.2$, the rupture propagates at approximately $1.55 C s$, (c) Schematic diagram showing the top view on the surface highlighting the location of the stations, as well as the arrival of the shear Mach front. The green triangles indicate the locations of the stations. The epicenter is marked by a yellow star. The transition point is marked by the green square and associated error bars. The green arrow indicates the rupture propagation direction.

We observe intense ground shaking associated with the arrival of the supershear Mach cone at the station and we identify this arrival by the red dashed line. Through measuring the change in ground motion associated with the supershear Mach front, we observe that the ratio of the fault parallel $\delta \dot{u}^s_{FP}$ to the fault normal component $\delta \dot{u}^s_{FN}$ is approximately $\sim 1.2$. As discussed by Mello et al. 2016, these changes correspond to the shear part of the velocity signal, and are due to the arrival of the shear Mach lines [7]. The ratio of the changes in the particle velocities has been theoretically shown by Mello et al. 2016 to depend uniquely on the ratio of the rupture speed and the shear wave speed as follows $\delta \dot{u}^s_{FP}/\delta \dot{u}^s_{FN} = \sqrt{(V_r/C_s)^2 - 1}$. This relationship is also shown schematically in Figure 2b. Accordingly, and as indicated in the figure, for a ratio of $1.2$, the corresponding supershear rupture speed is $\sim 1.5 C_s$.

Furthermore, in Figure 2a, the black dashed line indicates the eventual arrival of the trailing Rayleigh signature which represents the remnant of the initially sub-Rayleigh rupture before it transitioned to supershear. Figure 2c is a top view detailing the location of the three stations relative to the epicenter, highlighting the transition length $L_T$ after which the rupture speed $V_r$ exceeds the shear wave speed $C_s$. It also shows the shear Mach cone interaction with the stations.

Based on the geometry of Figure 2c, and assuming that the rupture tip initially propagates at $V_r = C_r$ prior to transition between $(0,0)$ and $(0, L_T)$ and then transition to $V_r = 1.55 C_s$ till it arrives at the twin stations at $x_2$, we can estimate a transition length $L_T$ by further assuming that the stations are located on the fault [8,13].

$$L_T = C_R \frac{x_2 - t_s V_r}{C_R - V_r} \qquad (1)$$

Where, $t_s$ is the arrival time of the shear Mach cone to the station which can be obtained from Figure 2a (red dashed line), and $V_r$ is the supershear rupture speed $1.55 C_s$. In the above relationship, $x_2$ is furnished as $\sqrt{L_2^2 - d^2}$ as shown in the insert of Figure 1, where $L_2$ is the distance of the twin stations from the hypocenter at depth $d$. $L_2$ is estimated based on the P-arrival time (first disturbance) from the hypocenter location to the station, and the assumed dilatational wave speed $C_p$ as we will describe shortly.

## Evidence of sub–Rayleigh to supershear transition in the TK:4165 station record

Similar to Figure 2a, Figure 3a shows the time histories of the particle velocities along the fault parallel, the fault normal, and the vertical directions obtained from station TK:4165 (AFAD) [2]. However, this record is qualitatively different from the record shown in Figure 2a. Indeed, we observe here that the fault normal velocity component is larger than the fault parallel component, which is characteristic of a primarily sub-Rayleigh rupture. However, careful examination of the fault parallel record indicates the presence of a small but well defined pulse ahead of the Rayleigh signature as indicated in the top panel of Figure 3a (shaded region). We believe that this feature is a supershear pulse, which has just been formed ahead of the primary rupture which is still propagating at the Rayleigh wave speed. Accordingly, we hypothesize that station TK:4165 is located very close to the point where the rupture transitioned from sub-Rayleigh to supershear. It should be noted that the probability of capturing the early stages of Rayleigh to supershear rupture transition is very low, and has never been observed before in a near fault field record. However, this transition has been reported experimentally in laboratory earthquakes performed by Rosakis et al 2004 [14] and Mello et al 2016 [7] (We refer the reader to Figure 14 in [7] for illustration). Specifically, Mello et al 2016 captured this transition by comparing dynamic, full field photoelastic images of the initial stages of the formation of the supershear pulse with near fault particle velocity records measured at a location close to the transitioning rupture and by further correlating the two measurement techniques. The velocity records were obtained experimentally by a pair of laser velocimeters recording the fault parallel and fault normal components [7].

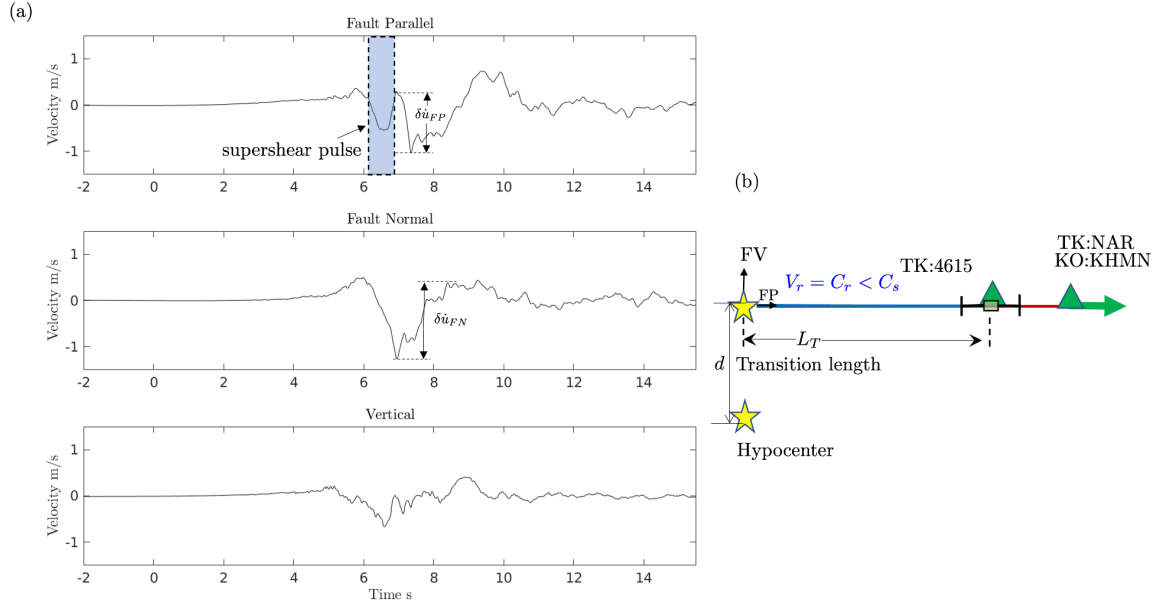

Figure 3. The transition from sub-Rayleigh to supershear rupture propagation is captured by the TK:4615 station. (a) The instrument corrected records of the fault parallel, fault normal, and vertical particle velocities. The highlighted region indicates the emergence of a supershear pulse ahead of the characteristic signature of a sub-Rayleigh rupture. (b) A schematic of the location of the station relative to the epicenter and hypocenter (yellow stars) location. The green triangle indicates the location of the stations. The epicenter is marked by a yellow star. The transition point is marked by the green square and associated error bars. The green arrow indicates the rupture propagation direction. Station TK:4615 is located within close proximity to the transition point.

To investigate the validity of this hypothesis, related to supershear transition and the location of TK:4615, we present a preliminary analysis by comparing the location of the station $x_1$ to our independent estimate of $L_T$ obtained from the twin stations record shown in Figure 2a. In order to do this, we assume $C_s = 3320$ m/s, and $C_p = 5780$ m/s which correspond to a Poisson's ratio of $0.25$, and are in good agreement with velocity models for the southern Turkey region [3]. It follows then that $C_R = 3050$ m/s and $V_r = 4960$ m/s. Based on the P-arrival time at the twin stations and using the above $C_p$ leads to $L_2 = 23.7$ km. We note that for a hypocenter depth of $d = 10.9$ km, equation (1) yields a transition length $L_T = 19.45$ km. We then use the P-wave arrival time at station TK:4165 to identify its distance from the hypocenter $L_1 = 22.3$. Using the Pythagorean theorem, we compute the epicentral distance of station TK:4165 as $x_1 = 19.45$ km. For this particular choice of depth $d$, we observe that the location of the station TK:4165 coincides with the location of the sub-Rayleigh to supershear transition, which is consistent with our hypothesis. This estimate of depth of $10.9$ km is within the range predicted by the different agencies (AFAD and USGS) [1,2]. Furthermore, computing the distance between the twin stations and TK:4165 yields $\delta x = x_2 - x_1 = 1.6$ km along the fault strike direction. Since the total distance between the twin stations and TK:4165 is $\sim 2$ km, based on their respective coordinates, this computed difference in their epicentral distances is a plausible estimate.

## Discussion:

Our analysis of three near-field velocity records of the $M_w$ 7.8 Kahramanmaraş earthquake suggests the rupture that propagated on the splay fault had transitioned from sub-Rayleigh to supershear speed ( $V_r \sim 1.55 C_s$) at an epicentral distance of approximately 19.45 km. The near-field records also captured, for the first time, the in–situ transition mechanism from sub-Rayleigh to supershear propagation and provided a detailed window into the structure of the near–fault particle motions in both the fault parallel and fault normal directions. Since Mach fronts attenuate only weakly with distance, this early supershear transition on the splay fault may have enabled strong dynamic stress transfer to the nearby East Anatolian Fault and contributed to the continued rupture propagation and triggered slip in both the North East and South West directions. Indeed, prior studies have suggested that supershear ruptures are more effective in jumping across fault stepovers [15] and activation of nearby faults [16-18]. The early supershear transition on the splay fault may have been favored by the regional stress state. Seismological studies [19] suggest that the splay fault exists in a N16.4°E compression regime ( $\sigma_1$) and it is under the N80.8°W extension regime ($\sigma_3$). The estimated strike of the splay fault N22°E thus makes it close to being perpendicular to the direction of the minimum principal stress which reduces the overall normal stress on the fault. This may significantly reduce the fault strength parameter S (e.g. S<1) [14,20] and favors transition to supershear rupture over shorter distances. Overall, we hope that further studies of the regional stress field and the structure of the ground motion records will reveal more details about the nature of this complex multi-segment rupture that led to such a large-scale human tragedy. Future detailed numerical simulations and analog experimental investigations are also needed to better constrain the dynamics of complex fault zones, like the East Anatolian Fault Zone, beyond what is available from historical records and regional scaling relations. This will help reduce the impact of future hazards and better inform preparedness efforts.

## Acknowledgement

A.J.R. acknowledges support by the Caltech/MCE Big Ideas Fund (BIF), as well as the Caltech Terrestrial Hazard Observation and Reporting Center (THOR). He would also like to acknowledge the support of NSF (Grant EAR-1651235 and EAR-1651235). A.E. acknowledge support from the Southern California Earthquake Center through a collaborative agreement between NSF. Grant Number: EAR0529922 and USGS. Grant Number: 07HQAG0008 and the National Science Foundation CAREER award No. 1753249 for modeling complex fault zone structures.

## References:

[1] U.S. Geological Survey, URL :https://earthquake.usgs.gov/earthquakes/eventpage/us6000jllz/origin/detail (accessed 02/13/2023 7:30 PM PST)

[2] Disaster and Emergency Management Authority. (1973). *Turkish National Strong Motion Network* [Data set]. Department of Earthquake, Disaster and Emergency Management Authority. https://doi.org/10.7914/SN/TK (accessed 02/13/2023 7:30 PM PST)

[3] Acarel, Diğdem, et al. "Seismotectonics of Malatya Fault, Eastern Turkey." *Open Geosciences* 11.1 (2019): 1098-1111.

[4] Kandilli Observatory And Earthquake Research Institute, Boğaziçi University. (1971). *Kandilli Observatory And Earthquake Research Institute (KOERI)* [Data set]. International Federation of Digital Seismograph Networks. https://doi.org/10.7914/SN/KO (accessed 02/13/2023 7:30 PM PST)

[5] M. Beyreuther, R. Barsch, L. Krischer, T. Megies, Y. Behr and J. Wassermann (2010) ObsPy: A Python Toolbox for Seismology SRL, 81(3), 530-533 DOI: 10.1785/gssrl.81.3.530

[6] Rosakis, A. J., O. Samudrala, and D. Coker. "Cracks faster than the shear wave speed." *Science* 284.5418 (1999): 1337-1340.

[7] Mello, Michael, Harsha S. Bhat, and Ares J. Rosakis. "Spatiotemporal properties of Sub-Rayleigh and supershear rupture velocity fields: Theory and experiments." *Journal of the Mechanics and Physics of Solids* 93 (2016): 153-181.

[8] Mello, M., et al. "Reproducing the supershear portion of the 2002 Denali earthquake rupture in laboratory." *Earth and Planetary Science Letters* 387 (2014): 89-96.

[9] Dunham, Eric M., and Ralph J. Archuleta. "Evidence for a supershear transient during the 2002 Denali fault earthquake." *Bulletin of the Seismological Society of America* 94.6B (2004): S256-S268.

[10] Bouchon, Michel, et al. "How fast is rupture during an earthquake? New insights from the 1999 Turkey earthquakes." *Geophysical Research Letters* 28.14 (2001): 2723-2726.

[11] Zeng, Hongyu, Shengji Wei, and Ares Rosakis. "A Travel‐Time Path Calibration Strategy for Back‐Projection of Large Earthquakes and Its Application and Validation Through the Segmented Super‐Shear Rupture Imaging of the 2002 Mw 7.9 Denali Earthquake." *Journal of Geophysical Research: Solid Earth* 127.6 (2022): e2022JB024359.

[12] Dunham, Eric M., and Harsha S. Bhat. "Attenuation of radiated ground motion and stresses from three‐dimensional supershear ruptures." *Journal of Geophysical Research: Solid Earth* 113.B8 (2008).

[13] Rubino, V., A. J. Rosakis, and Nadia Lapusta. "Spatiotemporal properties of sub‐Rayleigh and supershear ruptures inferred from full‐field dynamic imaging of laboratory experiments." *Journal of Geophysical Research: Solid Earth* 125.2 (2020): e2019JB018922.

[14] Xia, Kaiwen, Ares J. Rosakis, and Hiroo Kanamori. "Laboratory earthquakes: The sub-Rayleigh-to-supershear rupture transition." *Science* 303.5665 (2004): 1859-1861

[15] Harris, Ruth A., and Steven M. Day. "Dynamics of fault interaction: Parallel strike‐slip faults." *Journal of Geophysical Research: Solid Earth* 98.B3 (1993): 4461-4472.

[16] Templeton, Elizabeth L., et al. "Finite element simulations of dynamic shear rupture experiments and dynamic path selection along kinked and branched faults." *Journal of Geophysical Research: Solid Earth* 114.B8 (2009).

[17] Rousseau, Carl-Ernst, and Ares J. Rosakis. "Dynamic path selection along branched faults: Experiments involving sub-Rayleigh and supershear ruptures." *Journal of Geophysical Research: Solid Earth* 114.B8 (2009).

[18] Bhat, Harsha S., et al. "Dynamic slip transfer from the Denali to Totschunda faults, Alaska: Testing theory for fault branching." *Bulletin of the Seismological Society of America* 94.6B (2004): S202-S213.

[19] Kartal R. F., Kadirioğlu F. T., Zünbül, S, (2013), Kinematic of east Anatolian fault and Dead sea fault, Conference Paper, October 2013, https://www.researchgate.net /publication/271852091.

[20] D. J. Andrews, Rupture velocity of plane strain shear cracks. *J. Geophys. Res.* 81, 5679–5687 (1976).